\documentclass{article}
\usepackage{amssymb}
\usepackage{amsfonts}
\usepackage{amsmath}
\usepackage{graphicx}
\begin{document}

\title{Network extreme eigenvalue : From mutimodal to scale-free networks}
\author{N. N. Chung$^1$, L. Y. Chew$^2$ and C. H. Lai$^{3,4}$ \\
\\
$^{1}$ Temasek Laboratories, \\
National University of Singapore,\\
 Singapore 117508 \\
$^{2}$ Division of Physics \& Applied Physics, \\
School of Physical \& Mathematical Sciences, \\
Nanyang Technological University,\\
21 Nanyang Link, Singapore 637371\\
$^3$Beijing-Hong Kong-Singapore Joint Centre \\
for Nonlinear and Complex Systems (Singapore),\\
National University of Singapore, \\
 Kent Ridge 119260, Singapore \\
$^4$Department of Physics, \\
National University of Singapore, \\
Singapore 117542
}
\maketitle

\begin{abstract}
The extreme eigenvalues of adjacency matrices are important indicators on the influences of topological structures to collective dynamical behavior of complex networks. Recent findings on the ensemble averageability of the extreme eigenvalue further authenticate its sensibility in the study of network dynamics. Here we determine the ensemble average of the extreme eigenvalue and characterize the deviation across the ensemble through the discrete form of random scale-free network. Remarkably, the analytical approximation derived from the discrete form shows significant improvement over the previous results. This has also led us to the same conclusion as [Phys. Rev. Lett. 98, 248701 (2007)] that deviation in the reduced extreme eigenvalues vanishes as the network size grows.
\end{abstract}

{\bf Network extreme eigenvalues are succinct descriptors of the influence of the underlying topological structure of a complex network on its dynamics. This makes them important predictors of epidemic threshold of infectious diseases that propagate within real world complex network. Indeed, the recent demonstration that these eigenvalues are ensemble averageable has provided further support for this view. In this paper, we study into the ensemble averageability of extreme eigenvalues through a new perspective: the connection between multi-modal network and scale free network. The discrete nature of the multi-modal network has allowed us to arrive at an improved analytical expression of the extreme eigenvalue for scale-free networks. The extreme eigenvalues calculated from our analytical expression are found to closely correspond to those obtained numerically, thus making significant improvement over earlier versions.  The implication is a more accurate estimate of epidemic threshold, which is important for the elucidation of how vulnerable a particular network structure is to epidemic spreading. Our results are also applicable to the evaluation of strategies that aim to contain the spread of infectious diseases through the adjustment of the topology of network structures.} \\

Many concepts in network science have been well recognized as fundamental tools for exploring the dynamics of complex systems. In particular, scale-free networks are used widely to describe and model social, biological and economic systems \cite{Albert02,Barabasi99,Albert99,Redner98,Jeong00}. In an ensemble of scale-free networks, although the degree distribution of the nodes remains the same, the topological structure of each individual network can be diverse with different connections introduced between the nodes. Such structural diversity can lead to discrepancy in dynamics of the individual network. Since the structural influences on certain dynamical processes are governed by the extreme eigenvalues of the network adjacency matrices \cite{MacCluer00,Boguna02,Ott02,Boguna03,Restrepo06,Feng06,Restrepo08,Pomerance09}, deviations in the extreme eigenvalues in network ensembles are of increasing interest. Recently, it is found that the extreme eigenvalues of adjacency matrices, despite fluctuate widely in an ensemble of scale-free networks, are well characterized by the ensemble average after normalized by functions of the maximum degrees \cite{Kim07}. Specifically, it has been proven that the probability of having greatly deviated extreme eigenvalues in the ensemble diminishes as the size of the network increases. Considering the rich assortment of possible structural configurations for scale-free networks in an ensemble, this averageability is significant as it implies that dynamical processes which are governed by the extreme eigenvalues can be described simply using the ensemble average without the need of incorporating the connection details of the individual network. In particular, the average of network synchronization ability and epidemic spreading threshold are shown to be well approximated by functions of the ensemble average of the eigenvalues. Therefore, finding a way to determine the ensemble average of the extreme eigenvalues becomes crucial to uncover the topological influences of the network structure on a number of network dynamical processes.

To the best of our knowledge, the extreme eigenvalue of adjacency matrix of random, undirected scale-free network has been analytically approximated up to the second order correction as $\lambda_H^2 \approx k_H + k_H^{(1)}-1$ which gives better precision over the previous result $\lambda_H^2 \approx k_H$ \cite{Goh01,Farkas01,Dorogovtsev03,Chung03}. Note that $k_H$ is the largest degree of the network and $k_H^{(1)}$ denotes the average degree of the first nearest neighbors of node $H$. The probability distribution of the largest degree $P_d(k_H)$ is given by the Fr\'{e}chet distribution and the ensemble average of $k_H$ can be calculated from $P_d(k_H)$. However, as we shall show later, both of these approximation to $\lambda_H$ can be further improved.

In this paper, we investigate the extreme eigenvalue of undirected scale-free network through its discrete form, the multimodal network. For directed network, the extreme eigenvalue can be obtained from Refs. \cite{Restrepo07,Ott09}. Benefited from the mathematical properties of multimodal network that are more tractable, we found a way to analytically determine the ensemble average of the extreme eigenvalues while investigating the circumstances under which individual network can be better represented by the ensemble average. In addition, for bimodal networks which are shown to be more robust than the scale-free networks against both random and target removal of nodes \cite{Valente04,Paul04}, we study the difference between them and scale-free networks in terms of the ensemble average of the extreme eigenvalue.

A multimodal network \cite{Tanizawa06} with $m$ modes contains $m$ distinct peaks in the degree distribution: $P(k) = \sum_{i=1}^m r_i \delta(k-k_i)$. Note that $\delta(x)$ is the Dirac's delta function, $r_i=r_1 a^{-(i-1)} $ and
$k_i=k_1 b^{-(i-1)}$ for $i=1, 2,\cdots ,m$. It is assumed that $a>1$ and $0<b<1$ such that the degree distribution of the multimodal network follows a power law
$P(k_i) = r_i \propto k_i^{-\beta} $,
and hence $r_1> r_2 > \cdots > r_m$ for $k_1<k_2< \cdots < k_m$. As $m \rightarrow \infty$, the multimodal network converges to a scale-free network.
The largest degree of the network is $k_m$, the smallest degree is $k_1$ which is between $1$ and $\langle k \rangle$, and we have
$b=\left(k_1/k_m\right)^{\frac{1}{m-1}} $.
The rest of the parameters can be determined through the following equations:
\begin{eqnarray}
\sum_{i=1}^m r_i &=& r_1 \sum_{i=1}^m a^{-(i-1)} =1 \,, \label{normalization}\\
\sum_{i=1}^m k_i r_i &=& k_1 r_1 \sum_{i=1}^m (ab)^{-(i-1)} = \langle k \rangle\,. \label{average}
\end{eqnarray}

We shall follow the method outlined in \cite{Goh01} to find $\lambda_H$ of the multimodal network.
Let $G$ be a graph with vertex set $V(G)=\{v_1, v_2, \cdots, v_m\}$ and adjacency matrix $A$. For each positive integer $n$, the number of different $v_ j- v_i$ walk of length $n$ in $G$, denoted by $y_{j \rightarrow i}(n)$, is the $(j,i)$-entry in the matrix $A^n$.
Note that two $u-v$ walks $W=(u=u_0, u_1, \cdots, u_k=v)$ and $W'=(u=v_0, v_1, \cdots, v_l=v)$ in a graph are equal if $k=l$ and $u_i=v_i$ for all $i$ with $0\leq i \leq k$ \cite{Gary}. For example, $W_1 = (v_0, v_1, v_0, v_2)$ is different from $W_2 = (v_0, v_2, v_0, v_2)$ and both should be included in the calculation of $y_{0 \rightarrow 2}(3)$. In other words, $y_{j \rightarrow i}(n)$ is the total number of all possible walks of length $n$ from node $j$ to $i$, including those go backwards. Take a fully connected network with $3$ nodes as an example, there are $3$ walks of
length $3$ from node $1$ to node $2$, i.e. $1 \rightarrow 2 \rightarrow 1 \rightarrow 2$,
$1 \rightarrow 3 \rightarrow 1 \rightarrow 2$ and $1 \rightarrow 2 \rightarrow 3 \rightarrow 2$. This corresponds to $(A^3)_{12}=3$.
In the eigen-decomposition form, we have
$A^n = v D^n v'$,
where $v$ is the square matrix whose columns are the eigenvectors of $A$, $v'$ denotes the inverse of $v$ and $D$ is the diagonal matrix whose diagonal elements are the corresponding eigenvalues, i.e. $D_{ll}=\lambda_l$. Hence,
\begin{equation}
y_{j \rightarrow i}(n)=(A^n)_{ji} = \sum_l \lambda_l^n \, v_{j,l} \, v'_{l,i} \,. \label{sum_lambdaH}
\end{equation}
Note that Eq.(\ref{sum_lambdaH}) gives a summation over the $n$th power of the eigenvalues. When $n$ is sufficiently large, the $n$th power of $\lambda_H$ will be much larger than the $n$th power of the rest of the eigenvalues. Therefore, $y_{j \rightarrow i}(n)$ can be approximated in terms of only the maximum eigenvalue $\lambda_H$ as
\begin{equation}
y_{j \rightarrow i}(n) \approx \lambda_H^n v_{j,H} v'_{H,i} \,. \label{MaxEig}
\end{equation}
Now if we consider the number of walks of length $n+2$ which start and terminate at node H,
\begin{equation}
y_{H \rightarrow H}(n+2) = y_{H \rightarrow H}(n) \, y_{H \rightarrow H}(2) + \sum_{j \neq H} y_{H \rightarrow j}(n) y_{j \rightarrow H}(2) \,,
\end{equation}
then according to Eq. (\ref{MaxEig}),
\begin{equation}
\lambda_H^2 \approx  y_{H \rightarrow H}(2) + \sum_{j \neq H} \frac{y_{H \rightarrow j}(n) \, y_{j \rightarrow H}(2)}{y_{H \rightarrow H}(n)} \,. \label{lambdaH}
\end{equation}
The first term on the right hand side of Eq. (\ref{lambdaH}) corresponds to the number of the nearest neighbors of node $H$, i.e. the largest degree of the network $k_H$. In \cite{Goh01}, the second term on the right hand side of Eq. (\ref{lambdaH})  is shown numerically to be very small for scale-free networks and is hence neglected.
Since we are interested in finding a better approximation to the ensemble average of the maximum eigenvalues, we shall include the second term on the right hand side of Eq. (\ref{lambdaH}) in the calculation of $\lambda_H$ through a statistical approach.

Starting from node $H$, the total number of all possible walks of length $n$ to any node in the network is
\begin{equation}
y_H(n) = \sum_j y_{H \rightarrow j} (n) = \sum_j (A^n)_{Hj} \,.
\end{equation}
Since out of the $N \langle k \rangle$ total number of in-links and out-links, $k_j$ are directing into node $j$, hence, the fraction of these walks that end up at node $j$ is approximately $\frac{k_j}{N \langle k \rangle}$. Therefore,
\begin{equation}
y_{H \rightarrow j}(n)  \approx y_H(n) \, \frac{k_j}{N \langle k \rangle} \,,
\end{equation}
and
\begin{equation}
 y_{H \rightarrow H}(n) \approx y_H(n)  \frac{k_H}{N \langle k \rangle} \,.
\end{equation}
From node $j$, the number of possible one-step walk is equal to the number of neighbors of node $j$, i.e. $y_j (1) = k_j$. Similarly, from a neighbor of node $j$, says $j_1$, the number of one-step walk is equal to the number of neighbors of node $j_1$, i.e. $y_{j_1} (1) = k_{j_1}$. Walking two steps from node $j$ is the same as walking one step from the neighbor of node $j$, hence,
\begin{eqnarray}
\nonumber
y_j(2) &=& \sum_{q=1}^{k_j} y_{j_q} (1) \\
\nonumber
&=& \sum_{q=1}^{k_j} k_{j_q} \\
&=& k_j k_j^{(1)} \,,
\label{twostep}
\end{eqnarray}
where $k_j^{(1)} = \sum_{q=1}^{k_j} k_{j_q} / k_j $, is the average degree of the first nearest neighbors of node $j$. Since node $j$ is one of the neighbors of nodes $j_q$, thus among the two-step walks that begin from node $j$, all walks that go from node $j$, through its neighbors, and then go back to node $j$, is included. This means that backward walks are included in the calculation of $y_j(2)$ in Eq (\ref{twostep}).
Therefore,
\begin{equation}
y_{j \rightarrow H}(2) \approx k_j \, k_j^{(1)} \, \frac{k_H}{N \langle k \rangle} \,.\label{walk2}
\end{equation}
Since $k_j k_j^{(1)}$ can be small, the approximation in Eq. (\ref{walk2}) may not be precise for each $j$. Hence, the approximation is applicable only as an average over all nodes with degree $k_j$ instead of each individual case.

For multimodal scale-free network, there is a finite number, $m$ of distinct degrees $k_i$, each with probability $r_i$. Thus,
\begin{equation}
\lambda_H ^2 \approx k_m + \sum_{i=1}^m R_i k_i^2 \frac{k_i^{(1)}}{\langle k \rangle} \,, \label{lambda_H2}
\end{equation}
where
\begin{equation}
R_i= \left\{
\begin{array}{cc}
	r_i & \mbox{for} \,  \, \, 1<i<m-1 \,,\\
	r_i-1/N  & \mbox{for} \, \, \, i=m\, .
\end{array}  \right.
\end{equation}
Equation (\ref{lambda_H2}) implies that $ \lambda_H $ depends on the specific way the nodes within the network are connected, which can deviate broadly across the ensemble. When the exponent, $\beta$ of a scale-free network is small, we have a more heavy-tailed degree distribution. This results in a larger variation in the distribution of $k_i^{(1)}$ in the network ensemble. Hence, the values of $\lambda_H$ in the ensemble deviate more. In general, $k_m \propto \sqrt{\langle k\rangle N}$. For multimodal network with fixed $k_1$, the parameter $b$ and hence $k_i$ will be fixed. In order to have larger $\langle k \rangle$, the fraction of large-degree node has to be higher and the fraction of small-degree node has to be lower, this results in a more heavy-tailed distribution with the exponent $\beta$ to be smaller. In other words, $\beta \propto 1/\langle k \rangle$.
When the network size is larger, $\langle k \rangle$ has to be smaller for a fixed value of $k_m$, $\beta$ is hence larger and the probability of having larger-degree node drops rapidly.  Therefore, the values of $\lambda_H$ in an ensemble of multimodal networks deviate less as the networks become more sparse. In an ensemble of sparse networks, the individual network can thus be well represented by the ensemble average.

On the other hand, for a fixed value of $\langle k \rangle$, the degree distribution of multimodal networks vary with different values of $k_1$. Specifically, $\beta \propto k_1$. For two multimodal networks $A$ and $B$ of same size but different values of $k_1$, the fraction of large-degree node for the network with smaller $k_1$, says network $A$, has to be larger in order for it to have the same average degree with network $B$. Hence, $\beta_A < \beta_B$. The choice of different values of $k_1$ can thus lead to deviation in $\lambda_H$. Specifically, $k_1=1$ gives the extreme eigenvalue that is the largest, and $\lambda_H$ decreases as $k_1$ increases (see Fig. 1).

For a random network, the average of the sum of the nearest neighbor degree is  $z_2=G_0'(1)G_1'(1)$ \cite{Newman}.
Note that $G_0(x)=\sum_{k=0}^\infty p_k x^k$ is the generating function for the probability distribution of node degree while
$G_1(x)=\sum_{k=0}^\infty k p_k x^k / \langle k \rangle $
is the generating function for the distribution of the degree of the vertices which we arrive
at by following a randomly chosen edge. Hence, for a random multimodal network without any degree-degree correlation,
$k_i^{(1)} \approx \langle k^2 \rangle /\langle k \rangle$ and
\begin{eqnarray}
\nonumber
\langle \lambda_H \rangle &=& \sqrt{k_m + \sum_{i=1}^m R_i k_i^2 \frac{\langle k^2 \rangle}{\langle k \rangle^2}} \\
&=& \sqrt{k_m +  \frac{\langle k^2 \rangle^2}{\langle k \rangle^2} - \frac{k_m}{N}}\,. \label{lambdaavg}
\end{eqnarray}
Note that the second moment $\langle k^2 \rangle$ is a converging function of $m$, specifically,
\begin{equation}
\langle k^2 \rangle = r_1 k_1^2 \, \frac{1-(ab^2)^{-m}}{1-(ab^2)^{-1}} \,,
\end{equation}
with $(ab^2)^{-m} \rightarrow 0$ for $m \rightarrow \infty$.

With Eq. \ref{lambdaavg}, we study the dependence of the ensemble average of the extreme eigenvalues on the mode number $m$ for multimodal networks.
Here, we set $k_1=\langle k \rangle /2$ and $k_m =\sqrt{\langle k \rangle N}$. Figure 2 show the ensemble average of the extreme eigenvalues for multimodal network with $m$ modes.
$\langle \lambda_H \rangle$ is the highest for the bimodal network and it decreases gradually as $m$ increases and converges to a finite value.
Meanwhile, it can be seen from Fig. 2 that $\langle \lambda_H \rangle$ is larger for bimodal networks. Thus, although the tolerance against both random and targeted removal of node is optimal for bimodal network, epidemic spreading is less controllable.

\begin{center}
\begin{figure}
    \includegraphics[scale=0.6]{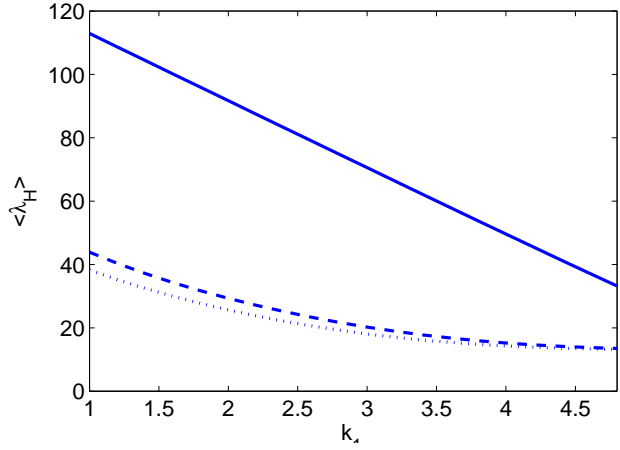}
    \caption{Dependence of $\langle \lambda_H \rangle$ on $k_1$ for the multimodal network with $\langle k \rangle =6$, $N=3 \times 10^3$ and $m=2$ (solid line), $10$ (dashed line) or $21$ (dotted line).}\label{fig1}
\end{figure}
\end{center}

\begin{center}
\begin{figure}
        \includegraphics[scale=0.6]{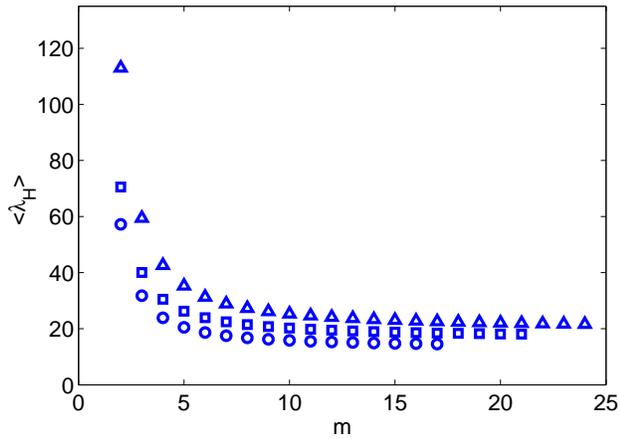}
        \caption{Dependence of $\langle \lambda_H \rangle$ on $m$ for the multimodal networks with $k_1=\langle k \rangle /2$. Note that sizes and average degrees of the networks are: (1) $\langle k \rangle =4, \, N=3000$ (circles), (2) $\langle k \rangle =6, \, N=3000$ (squares) and (3) $\langle k \rangle =6, \, N=8000$ (triangles).}\label{fig2}
\end{figure}
\end{center}

\begin{center}
    \begin{figure}
        \includegraphics[scale=0.5]{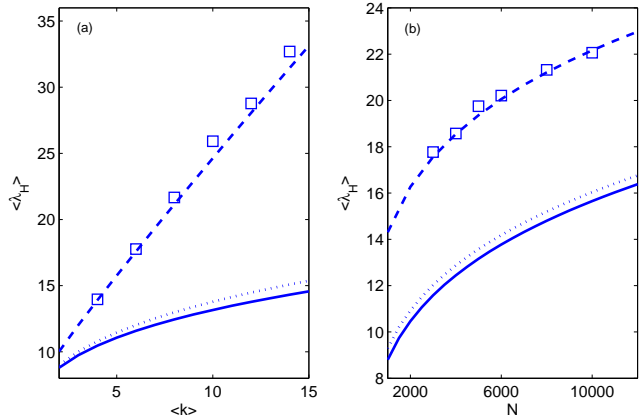}
        \caption{Dependence of $\langle \lambda_H \rangle$ on (a) $\langle k \rangle$ and (b) N  for ensembles of scale-free networks with $k_1=\langle k \rangle /2$. Note that $N=3 \times 10^3$ for (a) and $\langle k \rangle =6$ for (b). $\langle \lambda_H \rangle =\sqrt{k_H}$ are shown in solid curves, $\langle \lambda_H \rangle=\sqrt{ k_H + \langle k^2 \rangle / \langle k \rangle -1}$ are shown in dotted curves, analytical results from Eq. (\ref{lambdaavg}) are shown in dashed curves and numerical results of the BA model averaged over $200$ realizations of network are shown as squares.}\label{fig3}
    \end{figure}
\end{center}

To verify the accuracy of Eq. \ref{lambdaavg} , we compare analytical results of Eq. \ref{lambdaavg} to numerical results. For this, we generated a scale-free network with $k_1=\langle k \rangle /2$ and $k_m=\sqrt{\langle k \rangle N}$ using the Barab\'{a}si-Albert (BA) model \cite{Barabasi99}. An ensemble with randomized network topology is then created using the degree-preserving algorithm of Ref. \cite{Newman02}. For uncorrelated networks, we choose networks with assortativity coefficients near to zero. The maximum eigenvalue of each network is then computed and an ensemble average is obtained. In Figs. 3, we show the dependence of $\langle \lambda_H \rangle$ on $\langle k \rangle$ and $N$. Note that the numerical results are shown as squares. Next, we compute the ensemble average for multimodal networks with the same parameters using Eq. (\ref{lambdaavg}) by having $r_m \geq 1/N$ so that there is at least one node with degree $k_m$.
In addition, we compare our results with those given by $\langle \lambda_H \rangle=\sqrt{k_H}$ and $\langle \lambda_H \rangle=\sqrt{ k_H  + \langle k^2 \rangle / \langle k \rangle -1}$. As shown in Fig. 3,  our results give values of $\lambda_H$ that are closer to the numerical results compared to approximation from the previous results.

Many real-world networks are not uncorrelated, instead they show either assortative or disassortative mixing on their degree. For instance, the physics coauthorship network in Ref. \cite{Newman01} is assortative while the world-wide web network is disassortative \cite{Barabasi99}. For ensembles of network with identical degree distribution, $\langle \lambda_H \rangle$ of assortative networks with a preference of high-degree nodes to link to other high-degree nodes are larger than $\langle \lambda_H \rangle$ of disassortative networks. For these networks, $k_i^{(1)} \propto k_i^{-\nu}$ with $\nu>0 $ for disassortative networks and $\nu<0 $ for assortative network. Hence, although Eq. (\ref{lambdaavg}) gives ensemble average of $\lambda_H$ for randomly connected networks, it can be generalized as
\begin{equation}
\langle \lambda_H \rangle= \sqrt{k_m + \sum_{i=1}^m \frac{R_i k_i^{2-\nu}}{\langle k \rangle}} \,. \label{lambdaavgA}
\end{equation}
for correlated networks.
In fact, deviation in the extreme eigenvalue is larger in network ensemble with varying assortivities. Nonetheless, as shown in Ref. \cite{Kim07}, fluctuation in the normalized extreme eigenvalue diminishes as the network size grows. In Fig. 4,  we show the distribution of the normalized extreme eigenvalue $\lambda_H^N$ for $N=1000, 3000$ and $4000$. Note that a BA network is first generated, then an ensemble is generated by implementing $(\sum_i k_i)^2$ links rewirings following the constraints outlined in Ref. \cite{Newman02} and degree correlations in the networks are those generated due to these constraints \cite{Kim07}. As the network size grow, the distribution of $\lambda_H^N$ becomes more peaked and the standard deviation $\sigma^N$ decreases.

\begin{center}
\begin{figure}
        \includegraphics[scale=0.5]{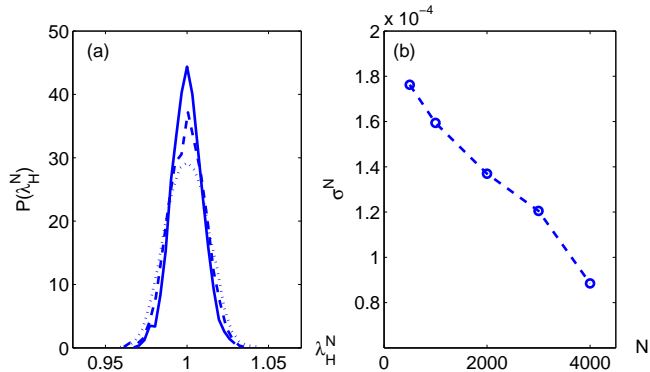}
        \caption{Numerical results for (a) the distribution $P(\lambda_H^N)$ of the normalized extreme eigenvalue $\lambda_H^N$ for $N=1000$ (dotted line), $3000$ (dashed line), $4000$ (solid line) and (b) the N dependence of the corresponding standard deviation $\sigma^N$. Note that all ensembles of network consist of $5000$ realizations of the networks with $\langle k \rangle =6$ and $k_1=3$.}\label{fig4}
\end{figure}
\end{center}

In conclusion, the ensemble averages of the extreme eigenvalues of scale-free networks can be determined more precisely through the multimodal networks with a large number of modes. Previous approximations give much lower values on ensemble average of the extreme eigenvalues, and this can cause an over-estimation of epidemic threshold. When dealing with network dynamics such as the epidemic spreading of the community-acquired meticilin-resistant Staphylococcus aureus (CA-MRSA) superbugs that are resistant to many antibiotics \cite{Kajita07}, over-estimating the epidemic threshold can lead to serious consequences. In view of this, the analytical solution derived from the multimodal network which is able to provide significantly closer approximation to the ensemble average of extreme eigenvalue of scale-free network is important. We have demonstrated that our analytical approximation predicted accurately the ensemble average of the extreme eigenvalues for scale free networks with $\beta \approx 3$ and $k_m=\sqrt{N \langle k \rangle}$. In fact, Eq. (\ref{lambdaavg}) is valid for a broad class of scale-free networks with different values of $\beta$ and $k_m$. While $k_m$ is a free parameter, the exponent $\beta$ can be adjusted by tuning parameters $a$ and $b$ through the relation: $\beta = - \ln a / \ln b +1$ \cite{Tanizawa06}. From Eq. (\ref{lambdaavg}), it is clear that $\langle \lambda_H \rangle$ increases with an increase of $\sqrt{k_m}$. Furthermore, it can be deduced from Eq. (\ref{lambdaavg}) that $\langle \lambda_H \rangle$ decreases as $\beta$ increases. This results from a drop in the variance in degree and the fraction of high degree nodes, as the exponent $\beta$ increases.

\bigskip

\noindent
{\bf \large{Acknowledgement} } \\

This work is supported by the Defense Science and Technology Agency of Singapore under project agreement of POD0613356.

\end{document}